%% file: main.tex
\documentclass[aps,prd,reprint,preprintnumbers,groupedaddress,nofootinbib,superscriptaddress]{revtex4-2}
\input{packages}
\usepackage{ATLAS_macros}

\begin{document}

\author{Benjamin Fuks\,\orcid{0000-0002-0041-0566}}
\email{fuks@lpthe.jussieu.fr}
\author{Mark D. Goodsell\,\orcid{0000-0002-6000-9467}}
\email{goodsell@lpthe.jussieu.fr}
\affiliation{Laboratoire de Physique Th\'{e}orique et Hautes \'{E}nergies (LPTHE),\\ UMR 7589, Sorbonne Universit\'{e} \& CNRS,\\ 4 place Jussieu, 75252 Paris Cedex 05, France\\[1ex]}
\author{Taylor Murphy\,\orcid{0000-0002-3215-9652}}
\email{murphy@lpthe.jussieu.fr}
\email{murphyt6@miamioh.edu}
\affiliation{Laboratoire de Physique Th\'{e}orique et Hautes \'{E}nergies (LPTHE),\\ UMR 7589, Sorbonne Universit\'{e} \& CNRS,\\ 4 place Jussieu, 75252 Paris Cedex 05, France\\[1ex]}
\affiliation{Department of Physics, Miami University\\
500 E. Spring St., Oxford, OH 45056, U.S.A.\\[1ex]}

\title{Monojets from compressed weak frustrated dark matter}

\input{Sections/abstract} 

\maketitle

\input{Sections/1_Intro}
\input{Sections/2_Model}
\input{Sections/3_DM}
\input{Sections/4_H152}
\input{Sections/5_LHC}
\input{Sections/6_Conclusion}

\acknowledgments

The work of B. F., M. D. G., and T. M. was supported in part by Grant ANR-21-CE31-0013, Project DMwithLLPatLHC, from the \emph{Agence Nationale de la Recherche} (ANR), France. 


\bibliographystyle{JHEP}
\bibliography{Bibliography/bibliography}

\end{document}

%% file: packages.tex
\usepackage[dvipsnames]{xcolor} 

\definecolor{mpl-orange}{HTML}{FFA500}
\definecolor{mpl-green}{HTML}{008000}

\usepackage{afterpage}
\usepackage{amsfonts, amssymb}
\usepackage{anyfontsize}
\usepackage{booktabs} 
\usepackage[hidelinks]{hyperref} 
\usepackage{makecell}
\usepackage{mathtools}
\usepackage{multirow}
\usepackage{physics}
\usepackage{slashed}
\usepackage[normalem]{ulem}
\usepackage{xspace} 
\usepackage[english]{babel}
\usepackage[section]{placeins}

\usepackage{newtxtext,newtxmath}

\newcommand{\ie}{\textit{i.e.}}

\def\MadGraph{{\sc MG5\_aMC}\xspace}
\newcommand{\madgraph}{{\sc MG5\_aMC}\xspace}
\newcommand{\delphes}{{\sc Del\-phes\,3}\xspace}
\newcommand{\madanalysis}{{\sc Mad\-A\-na\-ly\-sis\,5}\xspace}
\newcommand{\pythia}{{\sc Pythia\,8}\xspace}

\newcommand{\hackanalysis}{{\sc HackAnalysis}\xspace}
\newcommand{\BSMArt}{{\sc BSMArt}\xspace}

\def\ii{{\text{i}}}

\makeatletter
  \newcommand*{\transpose}{{\mathpalette\@transpose{}}}
  \newcommand*{\@transpose}[2]{\raisebox{\depth}{$\m@th#1\intercal$}}
\makeatother

\newcommand{\orcid}[1]{\begingroup
  \hypersetup{hidelinks}\href{https://orcid.org/#1}{\includegraphics[width=9pt]{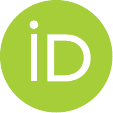}} \endgroup}

\makeatletter
\renewcommand{\fnum@figure}{\textsc{\figurename~\thefigure}} 
\makeatother

\makeatletter
\renewcommand*\l@subsubsection[2]{}
\makeatother

%% file: Sections/abstract.tex
\begin{abstract}

We extend the so-called hyperchargeless Higgs triplet model to include a weak triplet of Dirac fermions and a Dirac fermion $X$ transforming trivially under the Standard Model gauge group. We are motivated in part by a collection of anomalies that can be interpreted as a new scalar state with mass of approximately 152~GeV coupling to $W$ but not $Z$ bosons, which may be consistent with the electrically neutral triplet scalar in this model (provided that it mixes very slightly with the Standard Model Higgs boson). Meanwhile, the lightest neutral fermion in the model is stable and can be thermal dark matter with the correct relic abundance if it mixes lightly with the neutral triplet fermion, such that the dark matter is composed mostly of $X$. Because $X$ couples to the Standard Model only through a Yukawa-like interaction with the pair of triplets, this model falls into the frustrated dark matter paradigm. Finally, the spectrum of exotic fermions in this model can exhibit the strong compression favored by the current excess in the monojet channel, and evades multijet constraints in the region favored by monojets. In this work we explore this model's phenomenology and compare the parameter space regions best suited to each of the aforementioned excesses and constraints.

\end{abstract}

%% file: Sections/1_Intro.tex
\section{Introduction}
\label{s1}

Of the many searches for physics beyond the Standard Model (SM) during Run 2 of the LHC, several have reported excesses, although none of them are significant enough to exclude the SM. In recent work \cite{Agin:2024yfs}, we pointed out that four of these excesses were potentially overlapping. These corresponded to monojet searches in both the ATLAS and CMS experiments, and soft-lepton searches in both as well~\cite{ATLAS:2019lng,ATLAS:2021moa,CMS:2021edw}. In such cases, it is important to examine whether the observations can be explained by plausible models, in anticipation of more data.

In this paper, we consider the overlap of the monojet excesses with the possible signature of a Higgs-like particle with a mass around $152$~GeV \cite{Crivellin:2024uhc}. Phenomenological studies have suggested that this excess appears in new-physics analyses in the diphoton ($\gamma\gamma + X$) and $\gamma Z + X$ channels, and may furthermore be compatible with excesses revealed with multileptonic probes~\cite{vonBuddenbrock:2017gvy}. The $\gamma\gamma$ excess \cite{Crivellin:2021ubm} is interesting because the claimed significance is large---$4.3\sigma$ according to~\cite{Crivellin:2024uhc}---and yet has a potentially simple explanation in terms of a real $Y=0$ scalar field transforming as a triplet under the weak gauge group \cite{Ashanujjaman:2024pky,Crivellin:2024uhc}, among others \cite{Banik:2024ftv}. In this work, we point out that if we add electroweak fermions to this hyperchargeless Higgs triplet model \cite{Chabab:2018ert}, a construction first conceived for investigation at LEP \cite{Chardonnet:1993wd}, we obtain a prediction for excess monojet events that may at the same time also provide a viable dark matter (DM) candidate.

The DM is mostly aligned with a SM singlet fermion that couples indirectly to SM fields through a Yukawa-like coupling with one of the weak-triplet scalars and a triplet fermion. This coupling is the defining feature of \emph{frustrated dark matter} models \cite{Carpenter:2022lhj}, which have recently been explored with color-charged mediator fields. With mixing and coannihilation, some level of compression between the triplet and singlet fermions can produce the observed DM relic abundance through freeze out, and in the meantime the monojet excess seems to favor strong compression of just a few GeV. Thus we are presently interested in \emph{compressed} frustrated dark matter. The purpose of this work is to explore the overlap between the parameter spaces in this model favored by the (CMS) monojet analysis, searches for new physics in final states with multiple jets and missing energy ($E_{\text{T}}^{\text{miss}}$), the 152~GeV resonance explanation for various LHC anomalies, and the DM relic abundance.

This paper is organized as follows. Section \ref{s2} introduces our model's field content, interactions, and mass spectrum. In Section \ref{s3} we discuss the phenomenology of the singlet-like lightest neutral fermion, which can play the role of dark matter. Section \ref{s4} briefly notes how the triplet-like heavy scalar may appear as a $\gamma\gamma$ resonance consistent with recent observations. Finally, in Section \ref{s5}, we explore constraints on this model from the monojet channel and identify the region in which the monojet data favor our model over the Standard Model. We also discuss constraints from the multijet $+\, E_{\text{T}}^{\text{miss}}$ channel, which has some bins that overlap with the monojet analysis but is not considered to exhibit an excess. Section \ref{s6} concludes.

%% file: Sections/2_Model.tex
\section{Model discussion}
\label{s2}

In this section we introduce the model's field content and interactions. We take the opportunity to fix notation, describe the particle spectrum, and discuss the computer implementation of the model and the procedure utilized to generate benchmark points.

\subsection{Field content and mass spectrum}
\label{s2.1}

The field content of the new-physics sector of our model is summarized in Table \ref{tab:fields}. We begin with a $Y=0$ weak-triplet real scalar field $\Delta^{\transpose} = \sqrt{2}(\Delta^+, \Delta^0,\Delta^-)$. With the given normalization of $\Delta$, it is convenient to perform the unitary transformation
\begin{align}
    \Delta' = U\Delta\ \ \ \text{with}\ \ \ U = \frac{1}{\sqrt{2}}\begin{pmatrix}
        1 & 0 & 1\\
        \ii & 0 & -\ii\\
        0 & \sqrt{2} & 0
    \end{pmatrix}
\end{align}
so that $\Delta'$ (with the prime dropped henceforth) can be written as a $2\times 2$ matrix after projection onto the generators of the fundamental representation of $\mathrm{SU}(2)$. To wit:
\begin{align}
  \Delta_{ij} = \frac{1}{2}\sigma^A_{ij} \Delta^A = \frac{1}{\sqrt{2}} \begin{pmatrix}
        \Delta^0 & \sqrt{2}\Delta^+\\
        \sqrt{2}\Delta^- & -\Delta^0
    \end{pmatrix}.
\end{align}
Because $\Delta$ is real, the electrically charged scalars are each other's conjugates: $\Delta^+ = (\Delta^-)^{\dagger}$. We adopt an analogous representation for a weak-triplet vectorlike Dirac fermion $T$, which is also hyperchargeless.

\renewcommand{\arraystretch}{1.2}
\begin{table}
\centering
    \begin{tabular}{l@{\hspace{3ex}}c@{\hspace{3ex}}c@{\hspace{3ex}}c@{\hspace{3ex}}c}
    \toprule
    \toprule
Field & $\mathrm{SU}(2)_{\text{L}} \times \mathrm{U}(1)_Y$ & Spin & Components & $\mathbb{Z}_2$\\
    \midrule
$\Delta$ & $(\boldsymbol{3},0)$ & 0 & $\Delta^0, \Delta^{\pm}$ & $+1$\\
$T$ & $(\boldsymbol{3},0)$ & 1/2 & $T^-,T^0,T^+$ & $-1$\\
$X$ & $(\boldsymbol{1},0)$ & 1/2 & -- & $-1$\\
    \bottomrule
    \bottomrule
    \end{tabular}
    \caption{\label{tab:fields}Beyond the Standard Model (BSM) fields in the model. Both spin-1/2 fields are Dirac fermions.}
\end{table}
\renewcommand{\arraystretch}{1.0}

The final piece of this model is a SM singlet Dirac fermion $X$. In addition to the Standard Model, the Lagrangian comprises the terms
\begin{multline}
    \mathcal{L} = \frac{1}{4} \tr\, (D_\mu\Delta^\dag D^\mu\Delta) + \bar{T}(\ii \slashed{D} - m_T)T + \bar{X}(\ii \slashed{\partial} - m_X)X\\ + \mathcal{L}_{\text{int}} - V(\Phi,\Delta),\hspace{3.5cm}
\end{multline}
where
\begin{align}\label{eq:fDMLag}
-\mathcal{L}_{\text{int}} = y_T \tr \bar{T}\Delta T + y_X \bar{X}\tr \Delta T + \text{H.c.},
\end{align}
and the scalar potential (including the terms depending on the SM Higgs field $\Phi$ for clarity) is given by
\begin{multline}\label{eq:Vpot}
    V(\Phi,\Delta) = \mu^2 |\Phi|^2 + \frac{1}{2}\lambda (\Phi^{\dagger}\Phi)^2\\ + \frac{1}{2} \mu_{\Delta}^2 \tr \Delta^2 + \frac{1}{4} \lambda_{\Delta} \tr \Delta^4\\ + \sqrt{2}\delta_{\Delta} \Phi^{\dagger}\Delta \Phi + \frac{1}{2}\kappa_{\Delta} \,\Phi^{\dagger}\Phi \tr \Delta^2.
\end{multline}
All traces $\tr$ are over $\mathrm{SU}(2)_{\text{L}}$ indices.\footnote{The contraction of $\mathrm{SU}(2)_{\text{L}}$ indices in $\tr \bar{T}\Delta T$ is necessarily antisymmetric.} The electrically neutral component $\Delta^0$ of the triplet scalar obtains a vacuum expectation value (VEV) $\langle \Delta^0 \rangle \equiv v_{\Delta}$ upon electroweak symmetry breaking; this field mixes with the SM field $\Phi^0$, whose VEV is $\langle \Phi^0 \rangle \equiv v/\sqrt{2}$. (We require $(v^2 + v_{\Delta}^2)^{1/2} \approx 246~\text{GeV}$.) The scalar spectrum of this model therefore consists of a charged scalar $H^{\pm}$ with mass
\begin{align}\label{eq:HpmMass}
    m_{H^{\pm}}^2 = \frac{1}{2}\delta_{\Delta}\frac{v^2}{v_{\Delta}} + 2 \delta_{\Delta}v_{\Delta}
\end{align}
and two electrically neutral scalars. The mass matrix of the latter can be written as
\begin{align}\label{eq:h0MassMatrix}
    M_h^2 = \begin{pmatrix}
        \lambda v^2 & v(\kappa_{\Delta}v_{\Delta}-\delta_{\Delta})\\
        v(\kappa_{\Delta}v_{\Delta}-\delta_{\Delta}) & \frac{1}{2} \delta_{\Delta}\frac{v^2}{v_{\Delta}} + \kappa_{\Delta}v_{\Delta}^2
    \end{pmatrix}
\end{align}
in the Higgs basis $\{\Phi^0,\Delta^0\}$, after the elimination of $\mu$ and $\mu_{\Delta}$ via the tadpole equations. The Higgs basis and physical basis $\{h,H\}$ are related by an angle $\alpha$ satisfying (by construction)
\begin{multline}
    \sin 2\alpha = \frac{2v(\kappa_{\Delta}v_{\Delta}-\delta_{\Delta})}{m_H^2 - m_h^2}\\ \text{and}\ \ \ \tan 2\alpha = \frac{2v(\kappa_{\Delta}v_{\Delta}-\delta_{\Delta})}{\frac{1}{2}\delta_{\Delta}\frac{v^2}{v_{\Delta}}+\kappa_{\Delta}v_{\Delta}^2 - \lambda v^2},
\end{multline}
and the physical masses can be expressed in the limit $v_{\Delta}/v \ll 1$ as
\begin{align}
\nonumber    m_h^2 &= \lambda v^2 - 2 \delta_{\Delta}v_{\Delta} + 4 (\kappa_{\Delta}-\lambda)v_{\Delta}^2\\
\text{and}\ \ \    m_H^2 &= m_{H^{\pm}}^2 + [\lambda_{\Delta}- 4(\kappa_{\Delta}-\lambda)]v_{\Delta}^2.
\end{align}
We require the mixing of the doublet and triplet to be small on phenomenological grounds and consider spectra in which the lighter state $h$ is SM like. In this small-mixing ($\alpha$) limit, the masses reduce to
\begin{align}
    m_h^2 \approx \lambda v^2\ \ \ \text{and}\ \ \ m_H^2 \approx \frac{1}{2}\delta_{\Delta}\frac{v^2}{v_{\Delta}} + \lambda_{\Delta}v_{\Delta}^2,
\end{align}
and by comparison with \eqref{eq:HpmMass} it is clear that the triplet-like scalar splitting obeys the approximate relation
\begin{align}
m_{H^{\pm}}^2 - m_H^2 &\approx (2\delta_{\Delta}-\lambda_{\Delta}v_{\Delta})v_{\Delta},
\end{align}
which is quite small when $v_{\Delta}$ and $\delta_{\Delta}$ are tiny. In this scenario, the heavy scalar $H$ couples to pairs of $W$ bosons with coupling
\begin{align}
\mathcal{L} \supset \frac{1}{2} g^2\,(v \sin \alpha + 4 v_{\Delta} \cos \alpha)\,H W_{\mu}^+ W^{-\mu},
\end{align}
$g$ denoting the $\mathrm{SU}(2)_{\text{L}}$ coupling constant, and decays preferentially to $W^+W^-$ with branching fractions exceeding 90\%, though in many cases (including the scenarios discussed in this work) at least one $W$ boson may be off shell.

Meanwhile, the spectrum of BSM fermionic states consists of two pairs of charged fermions $\chi^{\pm}_{1,2}$ and two Dirac electrically neutral states $\chi^0_{1,2}$. In order to ensure the stability of the lightest electrically neutral fermion $\chi^0_1$, so that it may perform the role of dark matter, we impose a $\mathbb{Z}_2$ symmetry under which only $X$ and $T$ are odd. At leading order, the neutral fermions have masses
\begin{align}
    m_{\chi^0_{1,2}} = \frac{1}{2}\left[m_T+m_X \mp \sqrt{(m_T-m_X)^2 + 4(y_X v_{\Delta})^2}\right],
\end{align}
and the charged states are degenerate with mass
\begin{align}
    |m_{\chi^{\pm}_{1,2}}| = m_T\sqrt{1-\frac{1}{2}\left(y_T\, \frac{v_{\Delta}}{m_T}\right)^2}.
\end{align}
In the small-$v_{\Delta}$ limit, $\chi^{\pm}_{1,2}$ are furthermore nearly degenerate with $T^0$, the neutral component of the weak-triplet fermion. Loop effects break these (pseudo-)degeneracies. In the parameter space we examine below, wherein the physical state $\chi^0_2$ is triplet like, the spectrum consists of a singlet-like $\chi^0_1$ somewhat lighter than the three fermions $\{\chi^{\pm}_1,\chi^0_2,\chi^{\pm}_2\}$, which we write in order of ascending mass and are split nearly symmetrically by 1~GeV or less.

\subsection{Implementation; preliminary constraints}
\label{s2.2}

We implement the model in version 4.15.2 of \textsc{SARAH}~\cite{Staub:2008uz,Staub:2013tta,Goodsell:2014bna,Goodsell:2017pdq},\footnote{Note the factor of $1/2$ in the SM Higgs quartic coupling $\lambda$ in \eqref{eq:Vpot}, which is reflected in the \textsc{SARAH} implementation. The factor of $\sqrt{2}$ in \eqref{eq:Vpot}, which eliminates such factors in \eqref{eq:h0MassMatrix}, etc., is also implemented.} which we use to generate Fortran code using routines from the \textsc{SPheno} library~\cite{Porod:2003um,Porod:2011nf}. Fermion masses are computed including one-loop corrections, while decays are handled at leading order; scalar masses are computed including two-loop corrections \cite{Goodsell:2014pla,Goodsell:2015ira,Braathen:2017izn}. Moreover, we generate leading-order (LO) model files compatible with \textsc{CalcHEP}~\cite{Belyaev:2012qa} (hence \textsc{MicrOMEGAs}~\cite{Belanger:2010pz, Belanger:2013oya, Alguero:2023zol}) and \textsc{MadGraph5\texttt{\textunderscore}aMC@NLO} (\MadGraph)~\cite{Alwall:2014hca,Degrande:2011ua, Darme:2023jdn}. 

We use \BSMArt \cite{Goodsell:2023iac}, which calls the above tools and others detailed below, to steer several scans of the model parameter space featuring interesting dark matter and LHC monojet phenomenology. The input parameters are discussed below. For all scans, before any event generation is performed, we impose a set of experimental constraints based variously on Higgs measurements, electroweak precision tests, searches for flavor-changing processes, and a range of direct searches for BSM phenomena. The non-trivial constraints we make sure to satisfy include the following:
\begin{itemize}
    \item The $W$ boson mass $m_W$ must lie within the range compatible with the current global average, \emph{either} including the CDF II measurement with conservative uncertainty~\cite{deBlas:2022hdk},
    \begin{align*}
        m_W^{\text{CDF}} = (80.413 \pm 0.015)~\text{GeV},
    \end{align*}
    \emph{or} excluding that measurement \cite{ParticleDataGroup:2022pth},
    \begin{align*}
        m_W^{\text{\sout{CDF}}} = (80.377 \pm 0.012)~\text{GeV},
    \end{align*}
    \emph{or} (for flexibility) the gap between these two bands. (Most of our points fall in the \sout{CDF} band.) Either condition strongly constrains the triplet VEV $v_{\Delta}$ to be of $\mathcal{O}(1)$~GeV.
    \item The upward shift $\delta a_{\mu}$ in the anomalous magnetic dipole moment of the muon must not exceed the difference between the 2021 world experimental average \cite{Muong-2:2023cdq} and the 2020 Muon $g\!-\!2$ Theory Initiative calculation \cite{AOYAMA20201}:
    \begin{align*}
        \delta a_{\mu} \leq (g-2)_{\mu}^{\text{exp}} - (g-2)_{\mu}^{\text{th}} = 2.49 \times 10^{-9}.
    \end{align*}
    More recent results are available, so this condition could be quantitatively disputed, but in our parameter space $\delta a_{\mu}$ is of $\mathcal{O}(10^{-14})$. The shift is always tiny because the weak triplets have no hypercharge and the only BSM couplings to muons are induced by doublet-triplet scalar mixing.
    \item The scalar potential must be bounded from below to ensure the stability of the electroweak vacuum, and the scalar couplings must be small enough that all $2\to 2$ scalar processes satisfy the optical theorem (perturbative unitarity). The latter condition is fulfilled easily in our parameter space \cite{Ashanujjaman:2023etj}; vacuum stability is slightly less trivial and requires \cite{Chabab:2018ert}
    \begin{align*}
        \lambda,\lambda_{\Delta}>0\ \ \ \text{and}\ \ \ \frac{\kappa_{\Delta}}{\sqrt{2}} + \sqrt{2 \lambda_{\Delta} \lambda} > 0.
    \end{align*}
    \item Both neutral scalars must accommodate a range of experimental results, most notably the branching fractions $h \to \{ZZ,\text{invisible}\}$ \cite{CMS:2018amk,ATLAS:2018bnv} and the fiducial rates \cite{CMS:2015mca}
    \begin{align*}
        \sigma(gg,b\bar{b} \to H) \times \text{BF}(H \to \tau^+\tau^-).
    \end{align*}
\end{itemize}
The calculation of $m_W$, the electroweak precision observables including $\delta a_{\mu}$, and the perturbative unitarity check are set up by \textsc{SARAH} \cite{Goodsell:2018tti}. A suite of flavor physics constraints are checked using the FlavorKit extension of \textsc{SARAH} \cite{Porod:2014xia}. \textsc{BSMArt} calls \textsc{HiggsTools}~\cite{Bahl:2022igd} to constrain the scalar sector\footnote{\textsc{SARAH} gives coupling ratios to {\textsc{HiggsTools} to compute the constraints on the SM-like Higgs, which effectively incorporates higher-order effects from SM particles only; in the version of this model without extra fermions, the two-loop contributions to the diphoton partial width were recently computed~\cite{Degrassi:2024qsf} and shown to be modest.}}, and we interface to \textsc{Vevacious++}~\cite{Camargo-Molina:2013qva} to check for vacuum stability. As alluded to above, and detailed below, we use \textsc{MicrOMEGAs} to calculate the dark matter relic abundance and confront limits from direct- and indirect-detection experiments. Finally, we call \textsc{SModelS}~\cite{Waltenberger:2016vxp, Alguero:2021dig, MahdiAltakach:2023bdn} to perform a fast check for experimental constraints originating from colliders. Points are kept for simulation only if the \textsc{HiggsTools} $p$-value exceeds $p = 0.05$, the vacuum is stable, the \textsc{SModelS} $r$-value is below $r = 1$, and all other constraints mentioned above are satisfied. After some preliminary scans, in order to directly target the relevant model parameter space and reliably satisfy the constraints discussed above, our large scans for this work rely on the following input parameters:
\begin{equation}
\begin{array}{rlrl}
    \lambda &\in [0.2,0.3],  &\lambda_{\Delta} &\in [0,3.5],\\
\nonumber    \delta_{\Delta} &\in [0,2]~\text{GeV},&   \kappa_{\Delta} &\in [0,3.5],\\
\nonumber    v_{\Delta} &\in [1,10]~\text{GeV},
&y_T &\in [0,1],\\
\nonumber    y_X &\in [0,1],& 
 m_X&\in [95,400]~\text{GeV}, \end{array}
\end{equation}
and $m_T \in [96,435]$~GeV such that $m_T - m_X\in [1,35]~\text{GeV}$. Given small mixing between $X$ and $T^0$, this requirement enforces a similar condition on $\Delta m(\chi^0_2,\chi^0_1)$, which is important for our LHC analysis in Section \ref{s5}. Note that we use the Lagrangian parameters for input rather than the masses and mixing angles; this naturally ensures that the parameter space of all points is physical and respects perturbative unitarity.

%% file: Sections/3_DM.tex
\section{Phenomenology of the dark matter \texorpdfstring{$\chi^0_1$}{X01}}
\label{s3}

\begin{figure}
    \centering
    \includegraphics[scale=1.2]{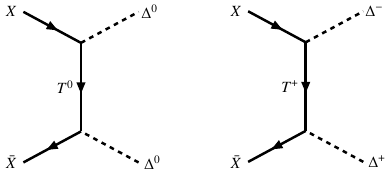}
    \caption{\label{fig:DManni1}Representative diagrams for $X$ pair annihilation. This singlet is pure frustrated dark matter, annihilating only to four-body SM final states, before it mixes with $\Delta^0$.}
\end{figure}

\begin{figure*}
    \centering
    \includegraphics[scale=1.2]{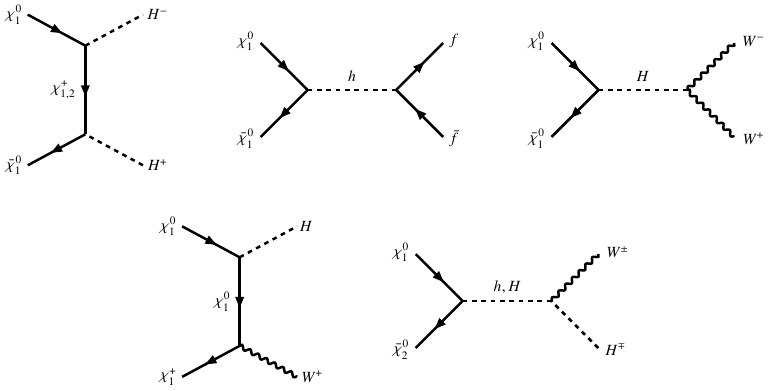}
    \caption{\label{fig:DManni2}Representative diagrams for (top) $\chi^0_1$ annihilation into a two-body SM final state and (bottom) coannihilation with $\chi^0_2,\chi^+_1$. $f$ denotes a SM fermion.}
\end{figure*}

The Yukawa-like coupling between the singlet fermion $X$ and the weak triplet pair, which appears in \eqref{eq:fDMLag} with coupling strength $y_X$, provides the only channels through which $X$ pairs can annihilate to visible particles at leading order. In terms of gauge eigenstates, depending on the hierarchy between the triplet scalar $\Delta$ and fermion $T$, the relevant processes are the $t$-channel annihilations $X\bar{X} \to \Delta^0 \Delta^0, \Delta^+\Delta^-$, with $\Delta^0$ subsequently decaying primarily to $W^+W^-$ and $\Delta^{\pm}$ primarily to $W^{\pm}Z$. Some representative diagrams are displayed in Figure \ref{fig:DManni1}. (While $X\bar{X} \to T^0\bar{T}^0$, etc., may be kinematically favored for some spectra, these processes cannot produce SM final states.)

This pattern, in which fermionic dark matter annihilates at tree level only to $\geq\! 4$ SM particles through a pair of mediating fields with identical gauge quantum numbers, is the hallmark of the frustrated dark matter paradigm \cite{Carpenter:2022lhj}. Typically, the DM candidate in frustrated DM models is strongly overabundant until its mass exceeds that of at least one mediator, at which point one or both mediators to which the DM annihilates can be approximately on shell. This parameter space permits relatively light $\mathcal{O}(100)$~GeV dark matter, and many of our benchmark points (see below) exhibit this kind of spectrum. But in the present model, as discussed in Section \ref{s2}, $X$ mixes with the electrically neutral component $T^0$ of the weak-triplet fermion. The neutral scalars $\Phi^0,\Delta^0$ also mix. Even when these mixings are small, so that the DM candidate $\chi^0_1$ is singlet like and the lightest CP-even scalar $h$ is SM like, they open additional annihilation channels such as $\chi^0_1 \bar{\chi}^0_1 \to W^+ W^-$ (both $s$- and $t$-channel diagrams contribute here) and $\chi^0_1 \bar{\chi}^0_1 \to q\bar{q}$ ($s$ channel, mediated by the SM-like scalar $h$). Furthermore, when the mass gap between $X$ and $T$ is small compared to their masses, coannihilation processes like $\chi^0_1 \bar{\chi}^0_2 \to H^+ W^-$, $\chi^0_1 \chi^{\pm}_1 \to W^{\pm}H$, $\chi^0_2 \chi^{\pm}_1 \to q\bar{q}'$ become efficient. Some representative diagrams for annihilation and coannihilation of various mass eigenstates are shown in Figure~\ref{fig:DManni2}. These two phenomena---mixing and compression---allow a singlet-like $\chi^0_1$ to freeze out with the relic abundance $\Omega h^2 = 0.12 \pm 0.001$ reported by the Planck Collaboration~\cite{Planck:2018vyg} even when it is lighter than the other BSM particles in the model.

In principle, both terrestrial (direct detection) and astrophysical (indirect detection) constraints apply to this realization of frustrated DM. In practice, however, DM-nucleon scattering has a small cross section in this model: it can proceed via SM-like Higgs exchange, as in the second diagram of Figure \ref{fig:DManni2}, but the rate is suppressed by both the SM Yukawa couplings and the small weak-triplet admixture of the 125~GeV scalar $h$. Thus it is understandable that \textsc{MicrOMEGAs} reports a trivial $p = 0.50$ for direct detection for all benchmark points. Constraints from searches for astrophysical $\gamma$ rays, on the other hand, are not so trivial \emph{a priori}. There are multiple open channels of DM annihilation into two-body SM states, notably $b\bar{b}$ and $W^+W^-$, and even more channels into two-body BSM states subsequently decaying into multiple SM species, all of which may produce observable deviations in the continuous $\gamma$-ray spectra measured by telescopes like \emph{Fermi}-LAT and MAGIC.\footnote{Monochromatic $\gamma$-ray signatures will also be produced by direct $\chi^0_1 \bar{\chi}^0_1 \to \gamma \gamma$ annihilation at one-loop order, but we relegate this calculation to future work.} To perform a rudimentary check of indirect-detection constraints on the points generated by our scans, we integrate the indirect-detection capabilities of \textsc{MicrOMEGAs}~\cite{Belanger:2010gh} with \textsc{BSMArt} for the first time. For each generated model point, we compare the annihilation cross section for every two-body SM final state to scraped limits from the 2016 joint \emph{Fermi}-LAT/MAGIC analysis \cite{MAGIC:2016xys} for the appropriate DM mass $m_{\chi^0_1}$. While these model-independent limits provide an incomplete picture, we expect limits on $2 \to 2 \to 4$ annihilation processes (with more than two SM particles) to be weaker since the resulting $\gamma$ rays should be softer. We also find that, for benchmarks predicting the correct DM relic abundance, the typical \emph{total} $\chi^0_1$ annihilation rates are between one and a few orders of magnitude smaller than the $\mathcal{O}(10^{-26})\ \text{cm}^3\,\text{s}^{-1}$ cross sections probed by the aforementioned experiments. (This tends to happen because, as mentioned above, coannihilations are required to achieve the correct relic abundance.) Thus we have reason to believe that our benchmark points are not excluded by indirect searches for dark matter.

As mentioned in Section \ref{s2}, we perform several scans for this work. We are naturally interested in obtaining points that predict an approximately correct DM relic abundance through freeze out; by ``approximately'' we mean within 20\% of the central \emph{Planck} value, in order to account for uncertain higher-order effects. Some of our scans impose this constraint. On the other hand, Sections \ref{s4} and \ref{s5} show results of scans that allow the dark matter relic abundance to vary freely; \ie, under- and overabundant dark matter are allowed in order to cover a wider parameter space and to better map the edge of the region accommodating the correct relic abundance. We clarify below when a plot shows only points with the approximately correct $\Omega h^2$. In both cases, we only retain model points that evade direct and indirect searches.

Upon investigation, we find---as hinted above---that many, if not most, scan points exhibiting the correct relic abundance achieve efficient DM annihilation either via $t$-channel processes mediated by charged fermions, such as $\chi^0_1 \bar{\chi}^0_1 \to H^+ H^-$ (which often supply a plurality of the total annihilation cross section); or through DM coannihilations with exotic charged fermions, such as $\chi^0_1 \chi^{\pm}_{1,2} \to W^{\pm} H$ or even $\chi^0_2 \bar{\chi}^0_2 \to W^+ W^-$. Such processes do not necessarily supply \emph{all} of the required annihilations, but they are seemingly always significant for benchmarks with efficiently annihilating dark matter. Both processes arise from the coupling between triplet scalar(s) and fermion(s) in \eqref{eq:fDMLag}, which defines a frustrated DM model. Thus the interplay between exotic fermions and scalars is critical for this model to include a viable dark matter candidate.

%% file: Sections/4_H152.tex
\section{Fitting the purported 152 GeV diphoton excess}
\label{s4}

A recent work \cite{Crivellin:2024uhc} examines potential evidence of Drell-Yan production of a $\sim\!\! 152$~GeV electrically neutral scalar in the sidebands of a Run 2 ATLAS search for new physics in final states including a SM-like Higgs boson \cite{ATLAS:2023omk}, and a concurrent ATLAS search for nonresonant Higgs pair production in leptonic or photonic final states~\cite{ATLAS:2024lhu}. The authors argue in~\cite{Crivellin:2024uhc} and other works \cite{Ashanujjaman:2024pky} that this targeted mass is motivated by excesses in a number of channels, including final states with multiple leptons resulting from $W$ boson decays \cite{Coloretti:2023wng}, the $b\bar{b} + E_{\text{T}}^{\text{miss}}$ channel \cite{ATLAS:2020fcp}, and several inclusive and exclusive $\gamma \gamma$ channels. In prior work \cite{Crivellin:2021ubm}, (some of) the same authors claim that the combination of these channels provides evidence for a 152~GeV scalar with global (local) significances of 5.1 (4.8) standard deviations relative to the background hypothesis. These significances have been challenged \cite{Fowlie:2021ldv} and may be as small as 4.1 (3.5)$\sigma$, but in any case seem to merit investigation.  

\begin{figure*}
\includegraphics[scale=0.6]{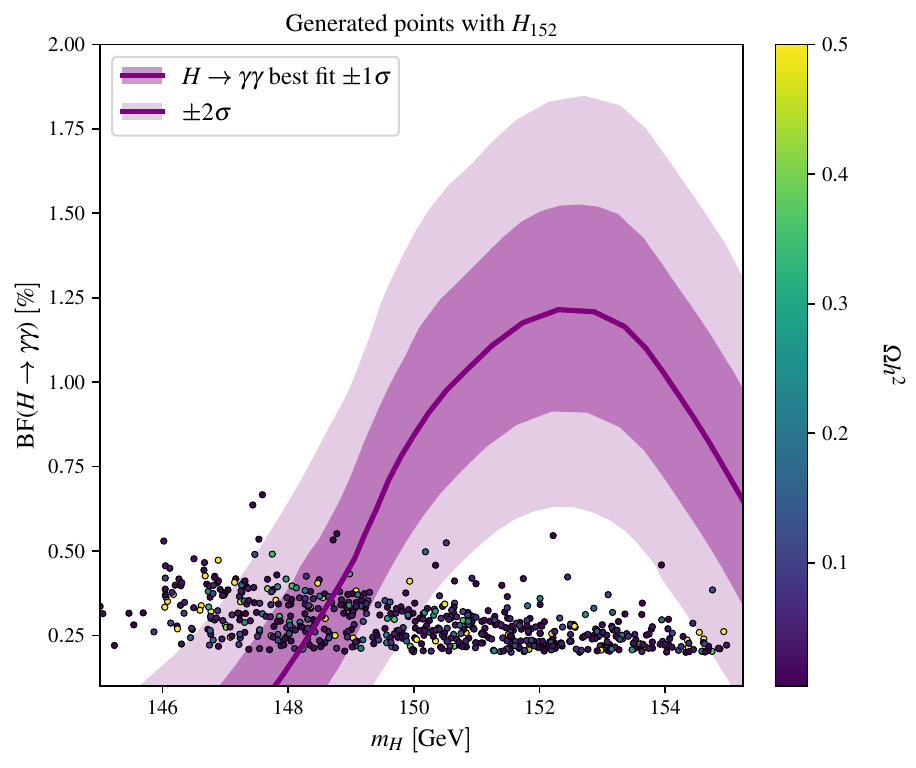}\hfill
\includegraphics[width=0.437\textwidth]{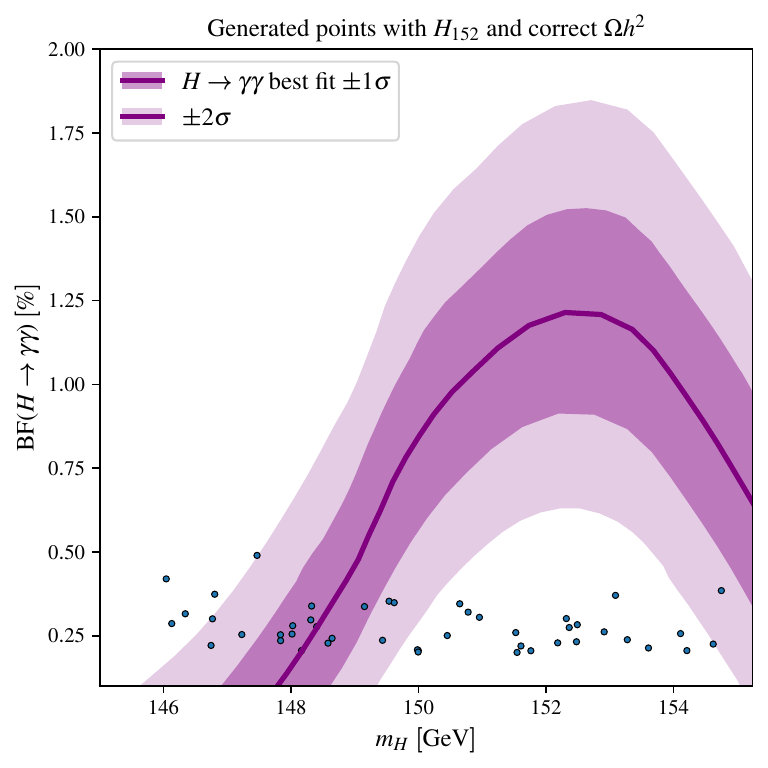}
\caption{\label{fig:H152scatter}Comparison of points used for monojet and multijet analyses to heavy scalar diphoton branching fraction favored by diphoton and multilepton anomalies, as computed by \cite{Crivellin:2024uhc}. Left panel shows all points and right panel keeps points with DM relic abundance $\Omega h^2$ within 20\% of the \emph{Planck} value.}
\end{figure*}

The authors of \cite{Crivellin:2024uhc} point out that the multilepton anomalies suggest a new particle with strong coupling to the $W$ boson, but that conversely the lack of an excess in the $ZZ \to 4\ell$ channel \cite{ATLAS:2020rej} points to a new particle with zero hypercharge. These ideas jointly motivate the authors to investigate the hyperchargeless Higgs triplet model. As mentioned above, they consider Drell-Yan production of the electrically neutral $H$ with its charged partner $H^{\pm}$, followed by the decays $H \to \gamma\gamma$ and $H^{\pm} \to \{WZ, tb, cs, \tau\nu\}$ (antiparticles depending on $H^+/H^-$). The authors identify eight signal regions in the ATLAS analyses \cite{ATLAS:2023omk,ATLAS:2024lhu} that are sensitive to these signals in principle, which for $m_H \approx 152$~GeV have a total Run 2 LHC cross section of $\sim\!\! 1$~pb. The analysis of this model is particularly convenient because (see Section \ref{s2}) the new neutral and charged scalars are nearly degenerate, and the branching fractions of $H^{\pm}$ can be expressed purely in terms of the neutral scalar's branching fraction to photons, $\text{BF}(H \to \gamma\gamma)$. Therefore in \cite{Crivellin:2024uhc}, the location and significance of the purported excess, within the framework of this model, is simply mapped onto the $(m_H, \text{BF}(H \to \gamma\gamma))$ plane. The authors find a $4.3 \sigma$ preference for $\text{BF}(H \to \gamma \gamma) \approx 1.2\%$ at $m_H \approx 152$~GeV. On the other hand, the authors also find tension in this model between the preferred value for $\text{BF}(H \to \gamma \gamma)$ and the SM-like Higgs diphoton branching fraction, $\text{BF}(h \to \gamma\gamma)$, and the requirements of vacuum stability and perturbative unitarity---so there is some question about the goodness of fit of this model to the 152~GeV excesses. Nevertheless, a $\text{BF}(H \to \gamma\gamma)$ fit within $3\sigma$ appears possible, so we use this new-physics scenario as inspiration and require $m_H \in [145,155]$~GeV. Since the addition of a fermionic multiplet with identical SM gauge quantum numbers does not abrogate the logic of \cite{Crivellin:2024uhc}, we take its results at face value.

The triplet-like neutral scalar $H$ couples weakly to photon pairs at one-loop order. The loops generating this coupling are composed of SM fermions (only top quarks are non-negligible), $W$ bosons, charged scalars $H^{\pm}$, and the charged fermions $\chi^{\pm}_{1,2}$. Most of these contributions have been reported previously \cite{Ashanujjaman:2023etj}; we include the $\chi^{\pm}$ loops for completeness. The diphoton decay rate of the heavy scalar $H$ can be expressed as
\begin{widetext}
    \begin{multline}\label{eq:diphotonRate}
        \Gamma(H \to \gamma \gamma) = \frac{G_{\text{F}} \alpha_{\text{EM}}^2}{128\sqrt{2} \pi^3}\,m_H^3\, \bigg| -\!\frac{4}{3} s_{\alpha} \mathcal{F}(\tau_t) + \frac{1}{\sqrt{2}}\,y_T v c_{\alpha} \left(\frac{1}{m_{\chi_1^{\pm}}} \mathcal{F}(\tau_{\chi^{\pm}_1}) - \frac{1}{m_{\chi_2^{\pm}}} \mathcal{F}(\tau_{\chi^{\pm}_2})\right) \\ -\frac{8 G_{\text{F}}}{2\sqrt{2}}\,v\left(\frac{1}{2} v s_{\alpha} + 2 v_{\Delta}c_{\alpha}\right)\mathcal{V}(\tau_W) - \frac{1}{2m_{H^{\pm}}^2}\, v (\lambda_{\Delta} v_{\Delta}c_{\alpha} - \kappa_{\Delta}  v s_{\alpha})\, \mathcal{S}(\tau_{H^{\pm}})\, \bigg|^2,
    \end{multline}
\end{widetext}
where $\tau_x = m_H^2/4m_x^2$ for some particle $x$ coupling directly to $H$, and
\begin{align}
\nonumber    \mathcal{F}(\tau) &= \frac{2}{\tau^2}\left[\tau+(\tau-1)f(\tau)\right],\\
\nonumber    \mathcal{V}(\tau) &= -\frac{1}{\tau^2}\left[2\tau^2 + 3\tau + 3(2\tau-1)f(\tau)\right],\\
    \mathcal{S}(\tau) &= -\frac{1}{\tau^2}\left[\tau-f(\tau)\right]
\end{align}
with
\begin{align}
    f(\tau) = \begin{cases}
        \arcsin^2 \sqrt{\tau}, & \tau \leq 1,\\
        -\dfrac{1}{4} \left(\ln \dfrac{1+\sqrt{1-\tau^{-1}}}{1-\sqrt{1-\tau^{-1}}} - \ii \pi \right)^2, & \tau > 1
    \end{cases}
\end{align}
are the usual one-loop three-point functions. The contribution from the charged fermions $\chi^{\pm}_{1,2}$ is quite small since the splitting between these species is loop induced and tiny. The largest contribution is from the $W$ boson loops but is strongly controlled by the doublet-triplet mixing: the prefactor of $\mathcal{V}(\tau_W)$ in \eqref{eq:diphotonRate} is suppressed either by $\sin \alpha$, which must be small to accommodate Higgs data, or by $v_{\Delta}$, which must be small for electroweak precision observables. Therefore the diphoton decay rate is altogether small---dwarfed in particular by the tree-level decay $\Gamma(H \to W^+ W^{-(*)})$---and the largest achievable diphoton branching fractions are of $\mathcal{O}(0.1)\%$.

The leading-order diphoton width \eqref{eq:diphotonRate} is automatically computed by \textsc{SARAH}/\textsc{SPheno} \cite{Staub:2016dxq} for each benchmark point generated by our scans in \textsc{BSMArt}. Figure \ref{fig:H152scatter} shows the results of a scan filtered to a relatively small collection of 618 points intended for LHC simulation, as discussed in Section \ref{s5} below. Our generated points are superimposed on the best-fit region scraped from Figure 5 of \cite{Crivellin:2024uhc} in the range $m_H \in [145,155.25]$~GeV. The left panel of our Figure \ref{fig:H152scatter} shows all of our points, which survive all constraints mentioned in Sections \ref{s2} and \ref{s3} but have unrestricted dark matter relic abundance. The right panel filters this collection once more to show only those points with $\Omega h^2 = 0.12 \pm 20\%$. Our large scan requires $\text{BF}(H \to \gamma \gamma) > 0.2\%$; hence the sharp cutoff visible in the left panel. In this model, it is not easy to generate points lying within two standard deviations of the central preferred value $\text{BF}(H \to \gamma \gamma) \approx 1.2\%$ for $m_H \approx 152$~GeV, though $3\sigma$ agreement is possible. This difficulty stems from the natural suppression of $\text{BF}(H \to \gamma\gamma)$ associated with small Higgs mixing angles. These findings are consistent with those shown in Figures 5 and 6 of \cite{Crivellin:2024uhc}.

%% file: Sections/5_LHC.tex
\section{Fitting the monojet excess; multijet~constraints?}
\label{s5}

\begin{figure}
    \centering
    \includegraphics[scale=1.2]{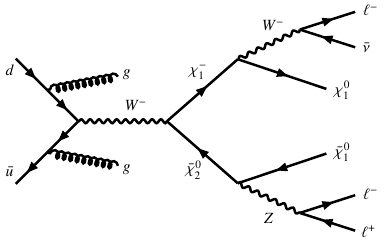}
    \caption{\label{fig:monojetSignal}Representative diagram for LHC charged fermion $\chi^{\pm}_1$ pair production with two hard jets. Processes of this type may satisfy the requirements of the monojet search CMS-EXO-20-004 if the fermions are compressed such that the SM decay products are soft.}
\end{figure}

The Run 2 CMS monojet search CMS-EXO-20-004 \cite{CMS:2021far} was originally interpreted within several frameworks, including simplified dark matter models and first-generation scalar leptoquarks, but applies in principle to fermion pair production in our model as displayed in Figure~\ref{fig:monojetSignal}. The dark matter $\chi^0_1$ can produce sizable $E_{\text{T}}^{\text{miss}}$ if the pair-produced fermions recoil off of hard initial-state radiation (ISR) jets. Vetoes of leptons and additional jets with $p_{\text{T}}$ of more than a few GeV optimize this analysis for particles decaying invisibly or softly, which can happen in models with small mass splittings. In our model, when the fermions are compressed, the decays of heavier species tend to proceed through off-shell weak bosons and result in soft leptons and hadrons. Interestingly, as we have explored in previous works \cite{Agin:2023yoq}, this analysis exhibits excesses in a variety of $E_{\text{T}}^{\text{miss}}$ bins. This analysis has been recast in \madanalysis \cite{Conte:2012fm, Conte:2014zja, Dumont:2014tja, DVN/IRF7ZL_2021} and adapted into \hackanalysis \cite{Goodsell:2021iwc,Goodsell:2024aig} for our convenience in previous work\footnote{We have checked that monojet predictions with \madanalysis, using the implementation provided by CMS~\cite{CMS:2021far}, and those obtained with our implementation in \hackanalysis agreed for all benchmarks in our scan on the basis of samples of $4\times 10^6$ events.} . We again use the latter implementation, which does not require a \delphes~\cite{deFavereau:2013fsa} simulation. We are also thus able to take advantage of a new feature of {\sc BSMArt} to run batches of events via gridpacks, and check after each run for convergence to a desired accuracy: this capability has been augmented by a check against the experimental sensitivity of each signal region, rather than converging when a given accuracy is reached. This saves significant processing time, since the signal regions for this analysis typically have very small efficiencies (since they are sliced into small bins in missing energy).

\begin{figure*}
    \centering
    \includegraphics[scale=0.75]{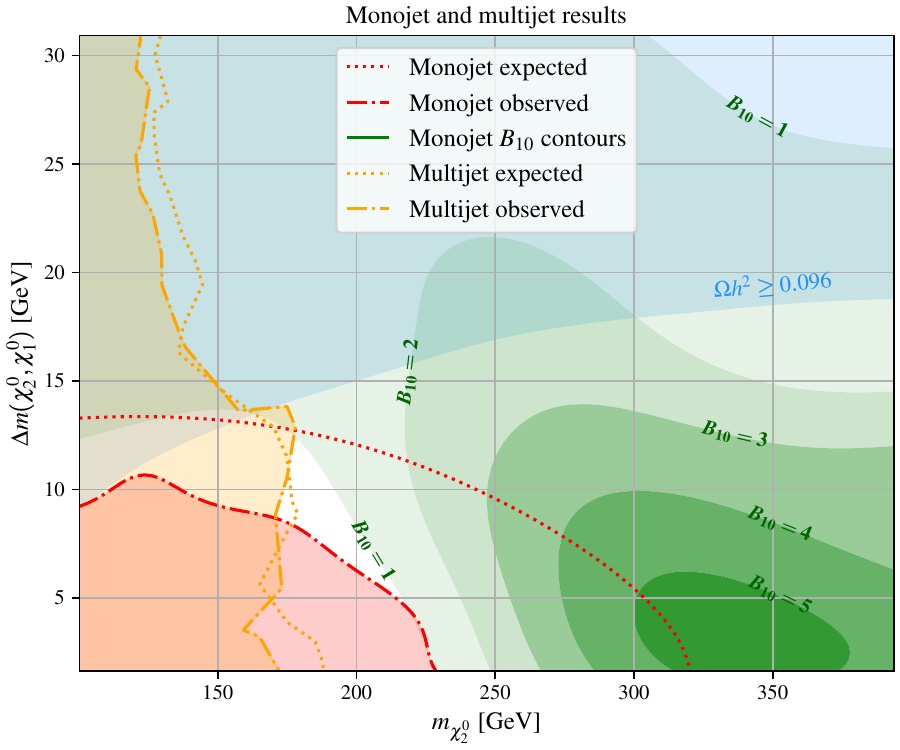}
    \caption{\label{fig:moneyplot}Summary of results from monojet (CMS-EXO-20-004) and multijet (CMS-SUS-19-006) recasts in our weak frustrated dark matter parameter space. Also shown for comparison is the approximate region in which the dark matter candidate $\chi^0_1$ can freeze out with relic abundance within 20\% of the observed value.}
\end{figure*}

We are also interested in whether our model is constrained by multijet data, since processes of the kind exemplified in Figure \ref{fig:monojetSignal} could populate some low-multiplicity (low-$N_{\text{jet}}$) bins of multijet analyses. The CMS Run 2 multijet search CMS-SUS-19-006 \cite{CMS:2019zmd} provides a good opportunity to compare multijet and monojet constraints, since these analyses use the same data. CMS-SUS-19-006 imposes cuts on $N_{\text{jet}}$, $N_{b\text{-jet}}$, and the scalar and vector sums of jet transverse momentum. It notably requires significant $E_{\text{T}}^{\text{miss}}$, such that it overlaps in principle with the monojet $(+\, E_{\text{T}}^{\text{miss}})$ analyses. It is particularly convenient to use this multijet analysis for the present exercise because it is already implemented in \madanalysis \cite{DVN/4DEJQM_2020}. Unlike the \textsc{HackAnalysis} monojet recast, this multijet recast requires \delphes for detector simulation.

As exemplified in Figure \ref{fig:monojetSignal}, the signal in our model relevant to CMS-EXO-20-004 is electroweak fermion pair production with a hard jet. We use \madgraph to simulate leading-order production of all six possible fermion pairs accompanied by up to two additional partons in the matrix element(s). The simulated processes include
\begin{align*}
    pp \to\ \chi^0_1 \chi^{\pm}_1,\ \chi^0_1 \chi^{\pm}_2,\ \chi^0_2 \chi^{\pm}_1,\ \chi^0_2 \chi^{\pm}_2,\ \chi^{\pm}_1 \chi^{\mp}_1,\ \text{and}\ \chi^{\pm}_2 \chi^{\mp}_2.
\end{align*} 
The triplet-like charged-current production processes $pp\to \chi_i^0\chi_j^\pm$, mediated by a $W^{\pm}$ boson, have the largest cross sections, followed by the neutral-current processes $pp\to \chi_i^\pm\chi_i^\mp$ at rates lower by a factor of roughly two. Parton showering and hadronization is performed using \pythia~\cite{Bierlich:2022pfr}. As mentioned above, for the \textsc{HackAnalysis} recast of CMS-EXO-20-004, we can dynamically generate events in batches until the (leading) efficiencies are under good statistical control. Nonetheless, owing to the tiny efficiencies in most signal regions of this analysis, for most parameter points we find that we still need to simulate our (self-imposed) upper limit of $10^7$ events. We separately simulate the same processes, with up to two additional hard jets, in order to use the \madanalysis recast of the multijet analysis CMS-SUS-19-006. In this case, just over $4 \times 10^6$ events are simulated for each model point. 

The results of our combined analysis are displayed in Figure~\ref{fig:moneyplot}. Here we show the expected and observed limits at 95\% confidence level (CL) \cite{Read:2002hq} for the recast monojet and multijet analyses in red and orange contours, respectively. Also shown for the monojet analysis are several isocontours of the \emph{Bayes factor} $B_{10}$ \cite{Fowlie:2024dgj}. For a new-physics search with phenomenological parameters $\theta$ and nuisances $\phi$, the Bayes factor is defined as
\begin{multline}\label{eq:B10Def}
    B_{10}(\theta) = \frac{Z(\theta)}{Z(\theta_0)}\\ \text{with}\ Z(\theta) = \int \text{d}\phi\, p(x\,|\,\theta,\phi)\,p(\phi\,|\,\theta),
\end{multline}
where $p(x\,|\,\theta,\phi)$ is the likelihood for some observation(s) $x$ and $p(\phi\,|\,\theta)$ is the prior for the nuisances. The Bayes factor gives the ratio of evidences $Z$ for a new-physics model with parameters $\theta$ to a model without new physics---for our purposes, the Standard Model---with parameters $\theta_0$. It therefore provides a quantitative measurement of whether a new model is favored or disfavored \emph{compared to} the SM, with values of $B_{10} > 1$ indicating that the evidence for the new model is stronger than that for the SM. When the nuisances are marginalized, the Bayes factor reduces to a likelihood ratio for the two hypotheses (new physics vs. SM). We calculate $B_{10}$ as a likelihood ratio for each of our frustrated DM benchmark points to evaluate whether our model is favored by the monojet search relative to the Standard Model. By plotting contours of $B_{10}$, we can find the region (if it exists) in which our model provides the best fit to the monojet excess compared to the scenario without any new physics.\footnote{Because this quantity provides an excellent visualization of parameter space regions that accommodate excesses better than the Standard Model, it would be a good choice for future analyses of exotic scalars in the vein of \cite{Crivellin:2024uhc}.}

As for other models we have explored in previous works, we find that the monojet analysis is expected (observed) to be sensitive only to points with $\Delta m(\chi^0_2,\chi^0_1) \lesssim 14$ (10)~GeV. On the other hand, the multijet constraints are much less sensitive to the mass splitting and are not too strong, hovering around $m_{\chi^0_2} \sim 125\text{--}175$~GeV. The shape of the excluded region is directly related to the signal cross section dependence, dominated by the associated production of a charged and a neutral new physics state. The signal efficiencies are indeed quite similar throughout the region of the parameter space considered. The significant excess in the monojet analysis is visible in the difference between the highest expected limit on $m_{\chi^0_2}$, around 320~GeV, and the observed limit at just above 225~GeV. The Bayes factor contours confirm that the monojet excess is associated with mass splittings of 6~GeV or less, and---interestingly, since the limit contours cannot give this information---that our weak frustrated DM model provides the best fit to the monojet excess, relative to the SM, for $m_{\chi^0_2} \sim 300\text{--}375$~GeV. In that part of the $(m_{\chi^0_2},\Delta m)$ plane, our model is at least five times more plausible than the SM. 

We finally mention the blue shaded region in approximately the upper half of Figure \ref{fig:moneyplot}. In this region, the relic abundance of the DM candidate $\chi^0_1$ can lie within 20\% of the observed value, $\Omega h^2 \approx 0.12$, for some choice(s) of input parameters. The edge of this region appears quite sharply in the large scan that we filtered in order to obtain the 618 points used for LHC simulation; while the high dimensionality of the input parameter space makes the ``correct $\Omega h^2$'' region quite diffuse, we find that beneath the edge that we have drawn in Figure \ref{fig:moneyplot}, the compression between $\chi^0_1$ and $\chi^0_2$ (hence $\chi^{\pm}_{1,2}$), the rate of coannihilations becomes too great such that the dark matter is underabundant in the present epoch. Unfortunately, the region with the approximately correct relic abundance only overlaps with the $B_{10}=2$ contour: our model does not seem to optimally fit the monojet excess and the preferred DM region. But it remains intriguing that there is some overlap.

%% file: Sections/6_Conclusion.tex
\section{Conclusions}
\label{s6}

In this work we have explored an extension of the hyperchargeless Higgs triplet model that includes additional electroweak fermions in a frustrated dark matter configuration. In addition to predicting a neutral scalar that couples to photon pairs at one-loop order, as suggested by the purported diphoton anomaly around 152~GeV, this model produces additional monojet events as suggested by the Run 2 excesses in that channel. We have performed a number of parameter space scans and used high-quality recasts of Run 2 CMS monojet and multijet analyses (CMS-EXO-20-004 and CMS-SUS-19-006) to explore all of this phenomenology and evaluate whether there is any overlap between the regions favored by any or all of the searches and measurements.

In addition to computing traditional limits at 95\% CL from the monojet and multijet analyses, we have computed the Bayes factor for the monojet analysis and plotted the isocontours of this quantity in order to visualize the region in which our model is specifically favored over the Standard Model by the monojet data. We find a region approximately centered on $(m_{\chi^0_2},\Delta m(\chi^0_2,\chi^0_1)) = (330,4)$~GeV in which our model is at least five times more plausible than the Standard Model with respect to that analysis. We also note some overlap between the region in which the lightest neutral fermion $\chi^0_1$ can freeze out with the observed dark matter relic abundance and an area in which our model provides a monojet fit at least twice as plausible as that offered by the SM. In the meantime, we have demonstrated that none of the parameter space in our model that is interesting for monojets is constrained by the multijet search. In the end, while there is no single point providing a perfect fit to all data, this model has multiple interesting characteristics and provides an excellent monojet signal.

These results suggest several interesting avenues of future work. First, a more thorough investigation of the dark matter phenomenology in this model is warranted, including a more accurate calculation of indirect-detection limits. Different variants of this model have been proposed to fit the CDF $W$ mass measurement \cite{Crivellin:2023xbu} and the 95~GeV excesses \cite{Ashanujjaman:2023etj}; it could be worthwhile to explore these alongside the monojet excess, but fitting the 95~GeV excesses is not straightforward since they appear in the $\gamma \gamma$, $b\bar{b}$, and $\tau^+ \tau^-$ channels in a pattern different from the branching fractions of the SM-like Higgs boson. Some more model building may be in order: it would be interesting to find a model with better overlap between the monojet-favored region and the area with correct DM relic abundance---and perhaps explain the mild excesses in the soft-lepton channels. The electroweak fermions in our model do not couple strongly enough to light leptons to produce more than a handful of events for $m_{\chi^0_2}$ of $\mathcal{O}(100)$~GeV, so a different scheme is needed to produce soft-lepton signals. Finally, on the recasting side, implementations of an ATLAS multijet search, the CMS Run 2 soft-lepton analysis, and an ATLAS search for new physics in final states with $b$-jets + $E_{\text{T}}^{\text{miss}}$ \cite{ATLAS:2021yij} are at various stages of development.

%% file: main.bbl
\providecommand{\href}[2]{#2}\begingroup\raggedright\begin{thebibliography}{10}

\bibitem{Agin:2024yfs}
D.~Agin, B.~Fuks, M.D.~Goodsell and T.~Murphy, \emph{{Seeking a coherent
  explanation of LHC excesses for compressed spectra}},
  \href{https://arxiv.org/abs/2404.12423}{{\ttfamily 2404.12423}}.

\bibitem{ATLAS:2019lng}
{\scshape ATLAS} collaboration, \emph{{Searches for electroweak production of
  supersymmetric particles with compressed mass spectra in $\sqrt{s}=$ 13 TeV
  $pp$ collisions with the ATLAS detector}},
  \href{https://doi.org/10.1103/PhysRevD.101.052005}{\emph{Phys. Rev. D}
  {\bfseries 101} (2020) 052005}
  [\href{https://arxiv.org/abs/1911.12606}{{\ttfamily 1911.12606}}].

\bibitem{ATLAS:2021moa}
{\scshape ATLAS} collaboration, \emph{{Search for
  chargino\textendash{}neutralino pair production in final states with three
  leptons and missing transverse momentum in $\sqrt{s} = 13$~TeV pp collisions
  with the ATLAS detector}},
  \href{https://doi.org/10.1140/epjc/s10052-021-09749-7}{\emph{Eur. Phys. J. C}
  {\bfseries 81} (2021) 1118}
  [\href{https://arxiv.org/abs/2106.01676}{{\ttfamily 2106.01676}}].

\bibitem{CMS:2021edw}
{\scshape CMS} collaboration, \emph{{Search for supersymmetry in final states
  with two or three soft leptons and missing transverse momentum in
  proton-proton collisions at $ \sqrt{s} $ = 13 TeV}},
  \href{https://doi.org/10.1007/JHEP04(2022)091}{\emph{J. High Energy Phys.}
  {\bfseries 04} (2022) 091}
  [\href{https://arxiv.org/abs/2111.06296}{{\ttfamily 2111.06296}}].

\bibitem{Crivellin:2024uhc}
A.~Crivellin, S.~Ashanujjaman, S.~Banik, G.~Coloretti, S.P.~Maharathy and
  B.~Mellado, \emph{{Growing Evidence for a Higgs Triplet}},
  \href{https://arxiv.org/abs/2404.14492}{{\ttfamily 2404.14492}}.

\bibitem{vonBuddenbrock:2017gvy}
S.~von Buddenbrock, A.S.~Cornell, A.~Fadol, M.~Kumar, B.~Mellado and X.~Ruan,
  \emph{{Multi-lepton signatures of additional scalar bosons beyond the
  Standard Model at the LHC}},
  \href{https://doi.org/10.1088/1361-6471/aae3d6}{\emph{J. Phys. G} {\bfseries
  45} (2018) 115003} [\href{https://arxiv.org/abs/1711.07874}{{\ttfamily
  1711.07874}}].

\bibitem{Crivellin:2021ubm}
A.~Crivellin, Y.~Fang, O.~Fischer, S.~Bhattacharya, M.~Kumar, E.~Malwa et~al.,
  \emph{{Accumulating evidence for the associated production of a new Higgs
  boson at the LHC}},
  \href{https://doi.org/10.1103/PhysRevD.108.115031}{\emph{Phys. Rev. D}
  {\bfseries 108} (2023) 115031}
  [\href{https://arxiv.org/abs/2109.02650}{{\ttfamily 2109.02650}}].

\bibitem{Ashanujjaman:2024pky}
S.~Ashanujjaman, S.~Banik, G.~Coloretti, A.~Crivellin, S.P.~Maharathy and
  B.~Mellado, \emph{{Explaining the $\gamma\gamma+X$ Excesses at $\approx$151.5
  GeV via the Drell-Yan Production of a Higgs Triplet}},
  \href{https://arxiv.org/abs/2402.00101}{{\ttfamily 2402.00101}}.

\bibitem{Banik:2024ftv}
S.~Banik and A.~Crivellin, \emph{{Explanation of the excesses in associated
  di-photon production at 152 GeV in 2HDM}},
  \href{https://arxiv.org/abs/2407.06267}{{\ttfamily 2407.06267}}.

\bibitem{Chabab:2018ert}
M.~Chabab, M.C.~Peyran\`ere and L.~Rahili, \emph{{Probing the Higgs sector of
  $Y=0$ Higgs Triplet Model at LHC}},
  \href{https://doi.org/10.1140/epjc/s10052-018-6339-2}{\emph{Eur. Phys. J. C}
  {\bfseries 78} (2018) 873}
  [\href{https://arxiv.org/abs/1805.00286}{{\ttfamily 1805.00286}}].

\bibitem{Chardonnet:1993wd}
P.~Chardonnet, P.~Salati and P.~Fayet, \emph{{Heavy triplet neutrinos as a new
  dark matter option}},
  \href{https://doi.org/10.1016/0550-3213(93)90101-T}{\emph{Nucl. Phys. B}
  {\bfseries 394} (1993) 35}.

\bibitem{Carpenter:2022lhj}
L.M.~Carpenter, T.~Murphy and T.M.P.~Tait, \emph{{Distinctive signals of
  frustrated dark matter}},
  \href{https://doi.org/10.1007/JHEP09(2022)175}{\emph{J. High Energy Phys.}
  {\bfseries 09} (2022) 175}
  [\href{https://arxiv.org/abs/2205.06824}{{\ttfamily 2205.06824}}].

\bibitem{Staub:2008uz}
F.~Staub, \emph{{SARAH}},  \href{https://arxiv.org/abs/0806.0538}{{\ttfamily
  0806.0538}}.

\bibitem{Staub:2013tta}
F.~Staub, \emph{{SARAH 4 : A tool for (not only SUSY) model builders}},
  \href{https://doi.org/10.1016/j.cpc.2014.02.018}{\emph{Comput. Phys. Commun.}
  {\bfseries 185} (2014) 1773}
  [\href{https://arxiv.org/abs/1309.7223}{{\ttfamily 1309.7223}}].

\bibitem{Goodsell:2014bna}
M.D.~Goodsell, K.~Nickel and F.~Staub, \emph{{Two-Loop Higgs mass calculations
  in supersymmetric models beyond the MSSM with SARAH and SPheno}},
  \href{https://doi.org/10.1140/epjc/s10052-014-3247-y}{\emph{Eur. Phys. J. C}
  {\bfseries 75} (2015) 32} [\href{https://arxiv.org/abs/1411.0675}{{\ttfamily
  1411.0675}}].

\bibitem{Goodsell:2017pdq}
M.D.~Goodsell, S.~Liebler and F.~Staub, \emph{{Generic calculation of two-body
  partial decay widths at the full one-loop level}},
  \href{https://doi.org/10.1140/epjc/s10052-017-5259-x}{\emph{Eur. Phys. J. C}
  {\bfseries 77} (2017) 758}
  [\href{https://arxiv.org/abs/1703.09237}{{\ttfamily 1703.09237}}].

\bibitem{Porod:2003um}
W.~Porod, \emph{{SPheno, a program for calculating supersymmetric spectra, SUSY
  particle decays and SUSY particle production at e+ e- colliders}},
  \href{https://doi.org/10.1016/S0010-4655(03)00222-4}{\emph{Comput. Phys.
  Commun.} {\bfseries 153} (2003) 275}
  [\href{https://arxiv.org/abs/hep-ph/0301101}{{\ttfamily hep-ph/0301101}}].

\bibitem{Porod:2011nf}
W.~Porod and F.~Staub, \emph{{SPheno 3.1: Extensions including flavour,
  CP-phases and models beyond the MSSM}},
  \href{https://doi.org/10.1016/j.cpc.2012.05.021}{\emph{Comput. Phys. Commun.}
  {\bfseries 183} (2012) 2458}
  [\href{https://arxiv.org/abs/1104.1573}{{\ttfamily 1104.1573}}].

\bibitem{Goodsell:2014pla}
M.D.~Goodsell, K.~Nickel and F.~Staub, \emph{{Two-loop corrections to the Higgs
  masses in the NMSSM}},
  \href{https://doi.org/10.1103/PhysRevD.91.035021}{\emph{Phys. Rev. D}
  {\bfseries 91} (2015) 035021}
  [\href{https://arxiv.org/abs/1411.4665}{{\ttfamily 1411.4665}}].

\bibitem{Goodsell:2015ira}
M.~Goodsell, K.~Nickel and F.~Staub, \emph{{Generic two-loop Higgs mass
  calculation from a diagrammatic approach}},
  \href{https://doi.org/10.1140/epjc/s10052-015-3494-6}{\emph{Eur. Phys. J. C}
  {\bfseries 75} (2015) 290}
  [\href{https://arxiv.org/abs/1503.03098}{{\ttfamily 1503.03098}}].

\bibitem{Braathen:2017izn}
J.~Braathen, M.D.~Goodsell and F.~Staub, \emph{{Supersymmetric and
  non-supersymmetric models without catastrophic Goldstone bosons}},
  \href{https://doi.org/10.1140/epjc/s10052-017-5303-x}{\emph{Eur. Phys. J. C}
  {\bfseries 77} (2017) 757}
  [\href{https://arxiv.org/abs/1706.05372}{{\ttfamily 1706.05372}}].

\bibitem{Belyaev:2012qa}
A.~Belyaev, N.D.~Christensen and A.~Pukhov, \emph{{CalcHEP 3.4 for collider
  physics within and beyond the Standard Model}},
  \href{https://doi.org/10.1016/j.cpc.2013.01.014}{\emph{Comput. Phys. Commun.}
  {\bfseries 184} (2013) 1729}
  [\href{https://arxiv.org/abs/1207.6082}{{\ttfamily 1207.6082}}].

\bibitem{Belanger:2010pz}
G.~Belanger, F.~Boudjema, A.~Pukhov and A.~Semenov, \emph{{micrOMEGAs: A Tool
  for dark matter studies}},
  \href{https://doi.org/10.1393/ncc/i2010-10591-3}{\emph{Nuovo Cim. C}
  {\bfseries 033N2} (2010) 111}
  [\href{https://arxiv.org/abs/1005.4133}{{\ttfamily 1005.4133}}].

\bibitem{Belanger:2013oya}
G.~Belanger, F.~Boudjema, A.~Pukhov and A.~Semenov, \emph{{micrOMEGAs\_3: A
  program for calculating dark matter observables}},
  \href{https://doi.org/10.1016/j.cpc.2013.10.016}{\emph{Comput. Phys. Commun.}
  {\bfseries 185} (2014) 960}
  [\href{https://arxiv.org/abs/1305.0237}{{\ttfamily 1305.0237}}].

\bibitem{Alguero:2023zol}
G.~Alguero, G.~Belanger, F.~Boudjema, S.~Chakraborti, A.~Goudelis, S.~Kraml
  et~al., \emph{{micrOMEGAs 6.0: N-component dark matter}},
  \href{https://doi.org/10.1016/j.cpc.2024.109133}{\emph{Comput. Phys. Commun.}
  {\bfseries 299} (2024) 109133}
  [\href{https://arxiv.org/abs/2312.14894}{{\ttfamily 2312.14894}}].

\bibitem{Alwall:2014hca}
J.~Alwall, R.~Frederix, S.~Frixione, V.~Hirschi, F.~Maltoni, O.~Mattelaer
  et~al., \emph{{The automated computation of tree-level and next-to-leading
  order differential cross sections, and their matching to parton shower
  simulations}}, \href{https://doi.org/10.1007/JHEP07(2014)079}{\emph{J. High
  Energy Phys.} {\bfseries 07} (2014) 079}
  [\href{https://arxiv.org/abs/1405.0301}{{\ttfamily 1405.0301}}].

\bibitem{Degrande:2011ua}
C.~Degrande, C.~Duhr, B.~Fuks, D.~Grellscheid, O.~Mattelaer and T.~Reiter,
  \emph{{UFO - The Universal FeynRules Output}},
  \href{https://doi.org/10.1016/j.cpc.2012.01.022}{\emph{Comput. Phys. Commun.}
  {\bfseries 183} (2012) 1201}
  [\href{https://arxiv.org/abs/1108.2040}{{\ttfamily 1108.2040}}].

\bibitem{Darme:2023jdn}
L.~Darm\'e et~al., \emph{{UFO 2.0: the \textquoteleft{}Universal Feynman
  Output\textquoteright{} format}},
  \href{https://doi.org/10.1140/epjc/s10052-023-11780-9}{\emph{Eur. Phys. J. C}
  {\bfseries 83} (2023) 631}
  [\href{https://arxiv.org/abs/2304.09883}{{\ttfamily 2304.09883}}].

\bibitem{Goodsell:2023iac}
M.D.~Goodsell and A.~Joury, \emph{{BSMArt: Simple and fast parameter space
  scans}}, \href{https://doi.org/10.1016/j.cpc.2023.109057}{\emph{Comput. Phys.
  Commun.} {\bfseries 297} (2024) 109057}
  [\href{https://arxiv.org/abs/2301.01154}{{\ttfamily 2301.01154}}].

\bibitem{deBlas:2022hdk}
J.~de~Blas, M.~Pierini, L.~Reina and L.~Silvestrini, \emph{{Impact of the
  Recent Measurements of the Top-Quark and W-Boson Masses on Electroweak
  Precision Fits}},
  \href{https://doi.org/10.1103/PhysRevLett.129.271801}{\emph{Phys. Rev. Lett.}
  {\bfseries 129} (2022) 271801}
  [\href{https://arxiv.org/abs/2204.04204}{{\ttfamily 2204.04204}}].

\bibitem{ParticleDataGroup:2022pth}
{\scshape Particle Data Group} collaboration, \emph{{Review of Particle
  Physics}}, \href{https://doi.org/10.1093/ptep/ptac097}{\emph{PTEP} {\bfseries
  2022} (2022) 083C01}.

\bibitem{Muong-2:2023cdq}
{\scshape Muon g-2} collaboration, \emph{{Measurement of the Positive Muon
  Anomalous Magnetic Moment to 0.20~ppm}},
  \href{https://doi.org/10.1103/PhysRevLett.131.161802}{\emph{Phys. Rev. Lett.}
  {\bfseries 131} (2023) 161802}
  [\href{https://arxiv.org/abs/2308.06230}{{\ttfamily 2308.06230}}].

\bibitem{AOYAMA20201}
T.~Aoyama, N.~Asmussen, M.~Benayoun, J.~Bijnens, T.~Blum, M.~Bruno et~al.,
  \emph{The anomalous magnetic moment of the muon in the standard model},
  \href{https://doi.org/https://doi.org/10.1016/j.physrep.2020.07.006}{\emph{Physics
  Reports} {\bfseries 887} (2020) 1}.

\bibitem{Ashanujjaman:2023etj}
S.~Ashanujjaman, S.~Banik, G.~Coloretti, A.~Crivellin, B.~Mellado and
  A.-T.~Mulaudzi, \emph{{SU(2)L triplet scalar as the origin of the 95~GeV
  excess?}}, \href{https://doi.org/10.1103/PhysRevD.108.L091704}{\emph{Phys.
  Rev. D} {\bfseries 108} (2023) L091704}
  [\href{https://arxiv.org/abs/2306.15722}{{\ttfamily 2306.15722}}].

\bibitem{CMS:2018amk}
{\scshape CMS} collaboration, \emph{{Search for a new scalar resonance decaying
  to a pair of Z bosons in proton-proton collisions at $\sqrt{s}=13 $ TeV}},
  \href{https://doi.org/10.1007/JHEP06(2018)127}{\emph{J. High Energy Phys.}
  {\bfseries 06} (2018) 127}
  [\href{https://arxiv.org/abs/1804.01939}{{\ttfamily 1804.01939}}].

\bibitem{ATLAS:2018bnv}
{\scshape ATLAS} collaboration, \emph{{Search for invisible Higgs boson decays
  in vector boson fusion at $\sqrt{s} = 13$ TeV with the ATLAS detector}},
  \href{https://doi.org/10.1016/j.physletb.2019.04.024}{\emph{Phys. Lett. B}
  {\bfseries 793} (2019) 499}
  [\href{https://arxiv.org/abs/1809.06682}{{\ttfamily 1809.06682}}].

\bibitem{CMS:2015mca}
{\scshape CMS} collaboration, \emph{{Search for additional neutral Higgs bosons
  decaying to a pair of tau leptons in $pp$ collisions at $\sqrt{s}$ = 7 and 8
  TeV}},  Tech. Rep. (2015).

\bibitem{Goodsell:2018tti}
M.D.~Goodsell and F.~Staub, \emph{{Unitarity constraints on general scalar
  couplings with SARAH}},
  \href{https://doi.org/10.1140/epjc/s10052-018-6127-z}{\emph{Eur. Phys. J. C}
  {\bfseries 78} (2018) 649}
  [\href{https://arxiv.org/abs/1805.07306}{{\ttfamily 1805.07306}}].

\bibitem{Porod:2014xia}
W.~Porod, F.~Staub and A.~Vicente, \emph{{A Flavor Kit for BSM models}},
  \href{https://doi.org/10.1140/epjc/s10052-014-2992-2}{\emph{Eur. Phys. J. C}
  {\bfseries 74} (2014) 2992}
  [\href{https://arxiv.org/abs/1405.1434}{{\ttfamily 1405.1434}}].

\bibitem{Bahl:2022igd}
H.~Bahl, T.~Biek\"otter, S.~Heinemeyer, C.~Li, S.~Paasch, G.~Weiglein et~al.,
  \emph{{HiggsTools: BSM scalar phenomenology with new versions of HiggsBounds
  and HiggsSignals}},
  \href{https://doi.org/10.1016/j.cpc.2023.108803}{\emph{Comput. Phys. Commun.}
  {\bfseries 291} (2023) 108803}
  [\href{https://arxiv.org/abs/2210.09332}{{\ttfamily 2210.09332}}].

\bibitem{Degrassi:2024qsf}
G.~Degrassi and P.~Slavich, \emph{{On the two-loop BSM corrections to
  $h\longrightarrow \gamma\gamma$ in a triplet extension of the SM}},
  \href{https://arxiv.org/abs/2407.18185}{{\ttfamily 2407.18185}}.

\bibitem{Camargo-Molina:2013qva}
J.E.~Camargo-Molina, B.~O'Leary, W.~Porod and F.~Staub,
  \emph{{$\mathbf{Vevacious}$: A Tool For Finding The Global Minima Of One-Loop
  Effective Potentials With Many Scalars}},
  \href{https://doi.org/10.1140/epjc/s10052-013-2588-2}{\emph{Eur. Phys. J. C}
  {\bfseries 73} (2013) 2588}
  [\href{https://arxiv.org/abs/1307.1477}{{\ttfamily 1307.1477}}].

\bibitem{Waltenberger:2016vxp}
{\scshape SModelS} collaboration, \emph{{SModelS: A Tool for Making Systematic
  Use of Simplified Models Results}},
  \href{https://doi.org/10.1088/1742-6596/762/1/012076}{\emph{J. Phys. Conf.
  Ser.} {\bfseries 762} (2016) 012076}.

\bibitem{Alguero:2021dig}
G.~Alguero, J.~Heisig, C.K.~Khosa, S.~Kraml, S.~Kulkarni, A.~Lessa et~al.,
  \emph{{Constraining new physics with SModelS version 2}},
  \href{https://doi.org/10.1007/JHEP08(2022)068}{\emph{J. High Energy Phys.}
  {\bfseries 08} (2022) 068}
  [\href{https://arxiv.org/abs/2112.00769}{{\ttfamily 2112.00769}}].

\bibitem{MahdiAltakach:2023bdn}
M.~Mahdi~Altakach, S.~Kraml, A.~Lessa, S.~Narasimha, T.~Pascal and
  W.~Waltenberger, \emph{{SModelS v2.3: enabling global likelihood analyses}},
  \href{https://doi.org/10.21468/SciPostPhys.15.5.185}{\emph{SciPost Phys.}
  {\bfseries 15} (2023) 185}
  [\href{https://arxiv.org/abs/2306.17676}{{\ttfamily 2306.17676}}].

\bibitem{Planck:2018vyg}
{\scshape Planck} collaboration, \emph{{Planck 2018 results. VI. Cosmological
  parameters}},
  \href{https://doi.org/10.1051/0004-6361/201833910}{\emph{Astron. Astrophys.}
  {\bfseries 641} (2020) A6}
  [\href{https://arxiv.org/abs/1807.06209}{{\ttfamily 1807.06209}}].

\bibitem{Belanger:2010gh}
G.~Belanger, F.~Boudjema, P.~Brun, A.~Pukhov, S.~Rosier-Lees, P.~Salati et~al.,
  \emph{{Indirect search for dark matter with micrOMEGAs2.4}},
  \href{https://doi.org/10.1016/j.cpc.2010.11.033}{\emph{Comput. Phys. Commun.}
  {\bfseries 182} (2011) 842}
  [\href{https://arxiv.org/abs/1004.1092}{{\ttfamily 1004.1092}}].

\bibitem{MAGIC:2016xys}
{\scshape MAGIC, Fermi-LAT} collaboration, \emph{{Limits to Dark Matter
  Annihilation Cross-Section from a Combined Analysis of MAGIC and Fermi-LAT
  Observations of Dwarf Satellite Galaxies}},
  \href{https://doi.org/10.1088/1475-7516/2016/02/039}{\emph{JCAP} {\bfseries
  02} (2016) 039} [\href{https://arxiv.org/abs/1601.06590}{{\ttfamily
  1601.06590}}].

\bibitem{ATLAS:2023omk}
{\scshape ATLAS} collaboration, \emph{{Model-independent search for the
  presence of new physics in events including $H\rightarrow\gamma\gamma$ with
  $\sqrt{s}$ = 13 TeV pp data recorded by the ATLAS detector at the LHC}},
  \href{https://doi.org/10.1007/JHEP07(2023)176}{\emph{J. High Energy Phys.}
  {\bfseries 07} (2023) 176}
  [\href{https://arxiv.org/abs/2301.10486}{{\ttfamily 2301.10486}}].

\bibitem{ATLAS:2024lhu}
{\scshape ATLAS} collaboration, \emph{{Search for non-resonant Higgs boson pair
  production in final states with leptons, taus, and photons in $pp$ collisions
  at $\sqrt{s}$ = 13 TeV with the ATLAS detector}},
  \href{https://doi.org/10.1007/JHEP08(2024)164}{\emph{JHEP} {\bfseries 08}
  (2024) 164} [\href{https://arxiv.org/abs/2405.20040}{{\ttfamily
  2405.20040}}].

\bibitem{Coloretti:2023wng}
G.~Coloretti, A.~Crivellin, S.~Bhattacharya and B.~Mellado, \emph{{Searching
  for low-mass resonances decaying into W bosons}},
  \href{https://doi.org/10.1103/PhysRevD.108.035026}{\emph{Phys. Rev. D}
  {\bfseries 108} (2023) 035026}
  [\href{https://arxiv.org/abs/2302.07276}{{\ttfamily 2302.07276}}].

\bibitem{ATLAS:2020fcp}
{\scshape ATLAS} collaboration, \emph{{Measurements of $WH$ and $ZH$ production
  in the $H \rightarrow b\bar{b}$ decay channel in $pp$ collisions at 13 TeV
  with the ATLAS detector}},
  \href{https://doi.org/10.1140/epjc/s10052-020-08677-2}{\emph{Eur. Phys. J. C}
  {\bfseries 81} (2021) 178}
  [\href{https://arxiv.org/abs/2007.02873}{{\ttfamily 2007.02873}}].

\bibitem{Fowlie:2021ldv}
A.~Fowlie, \emph{{Comment on \textquotedblleft{}Accumulating evidence for the
  associate production of a neutral scalar with mass around 151
  GeV\textquotedblright{}}},
  \href{https://doi.org/10.1016/j.physletb.2022.136936}{\emph{Phys. Lett. B}
  {\bfseries 827} (2022) 136936}
  [\href{https://arxiv.org/abs/2109.13426}{{\ttfamily 2109.13426}}].

\bibitem{ATLAS:2020rej}
{\scshape ATLAS} collaboration, \emph{{Higgs boson production cross-section
  measurements and their EFT interpretation in the $4\ell $ decay channel at
  $\sqrt{s}=$13 TeV with the ATLAS detector}},
  \href{https://doi.org/10.1140/epjc/s10052-020-8227-9}{\emph{Eur. Phys. J. C}
  {\bfseries 80} (2020) 957}
  [\href{https://arxiv.org/abs/2004.03447}{{\ttfamily 2004.03447}}].

\bibitem{Staub:2016dxq}
F.~Staub et~al., \emph{{Precision tools and models to narrow in on the 750 GeV
  diphoton resonance}},
  \href{https://doi.org/10.1140/epjc/s10052-016-4349-5}{\emph{Eur. Phys. J. C}
  {\bfseries 76} (2016) 516}
  [\href{https://arxiv.org/abs/1602.05581}{{\ttfamily 1602.05581}}].

\bibitem{CMS:2021far}
{\scshape CMS} collaboration, \emph{{Search for new particles in events with
  energetic jets and large missing transverse momentum in proton-proton
  collisions at $ \sqrt{s} $ = 13 TeV}},
  \href{https://doi.org/10.1007/JHEP11(2021)153}{\emph{J. High Energy Phys.}
  {\bfseries 11} (2021) 153}
  [\href{https://arxiv.org/abs/2107.13021}{{\ttfamily 2107.13021}}].

\bibitem{Agin:2023yoq}
D.~Agin, B.~Fuks, M.D.~Goodsell and T.~Murphy, \emph{{Monojets reveal
  overlapping excesses for light compressed higgsinos}},
  \href{https://doi.org/10.1016/j.physletb.2024.138597}{\emph{Phys. Lett. B}
  {\bfseries 853} (2024) 138597}
  [\href{https://arxiv.org/abs/2311.17149}{{\ttfamily 2311.17149}}].

\bibitem{Conte:2012fm}
E.~Conte, B.~Fuks and G.~Serret, \emph{{MadAnalysis 5, A User-Friendly
  Framework for Collider Phenomenology}},
  \href{https://doi.org/10.1016/j.cpc.2012.09.009}{\emph{Comput. Phys. Commun.}
  {\bfseries 184} (2013) 222}
  [\href{https://arxiv.org/abs/1206.1599}{{\ttfamily 1206.1599}}].

\bibitem{Conte:2014zja}
E.~Conte, B.~Dumont, B.~Fuks and C.~Wymant, \emph{{Designing and recasting LHC
  analyses with MadAnalysis 5}},
  \href{https://doi.org/10.1140/epjc/s10052-014-3103-0}{\emph{Eur. Phys. J.}
  {\bfseries C74} (2014) 3103}
  [\href{https://arxiv.org/abs/1405.3982}{{\ttfamily 1405.3982}}].

\bibitem{Dumont:2014tja}
B.~Dumont, B.~Fuks, S.~Kraml, S.~Bein, G.~Chalons, E.~Conte et~al.,
  \emph{{Toward a public analysis database for LHC new physics searches using
  MADANALYSIS 5}},
  \href{https://doi.org/10.1140/epjc/s10052-014-3242-3}{\emph{Eur. Phys. J.}
  {\bfseries C75} (2015) 56} [\href{https://arxiv.org/abs/1407.3278}{{\ttfamily
  1407.3278}}].

\bibitem{DVN/IRF7ZL_2021}
A.~Albert, \emph{{Implementation of a search for new phenomena in events
  featuring energetic jets and missing transverse energy (137 fb-1; 13 TeV;
  CMS-EXO-20-004)}},
  \href{https://doi.org/10.14428/DVN/IRF7ZL}{\emph{10.14428/DVN/IRF7ZL} (2021)
  }.

\bibitem{Goodsell:2021iwc}
M.D.~Goodsell and L.~Priya, \emph{{Long dead winos}},
  \href{https://doi.org/10.1140/epjc/s10052-022-10188-1}{\emph{Eur. Phys. J. C}
  {\bfseries 82} (2022) 235}
  [\href{https://arxiv.org/abs/2106.08815}{{\ttfamily 2106.08815}}].

\bibitem{Goodsell:2024aig}
M.D.~Goodsell, \emph{{HackAnalysis 2: A powerful and hackable recasting tool}},
   \href{https://arxiv.org/abs/2406.10042}{{\ttfamily 2406.10042}}.

\bibitem{deFavereau:2013fsa}
{\scshape DELPHES 3} collaboration, \emph{{DELPHES 3, A modular framework for
  fast simulation of a generic collider experiment}},
  \href{https://doi.org/10.1007/JHEP02(2014)057}{\emph{J. High Energy Phys.}
  {\bfseries 02} (2014) 057} [\href{https://arxiv.org/abs/1307.6346}{{\ttfamily
  1307.6346}}].

\bibitem{CMS:2019zmd}
{\scshape CMS} collaboration, \emph{{Search for supersymmetry in proton-proton
  collisions at 13 TeV in final states with jets and missing transverse
  momentum}}, \href{https://doi.org/10.1007/JHEP10(2019)244}{\emph{J. High
  Energy Phys.} {\bfseries 10} (2019) 244}
  [\href{https://arxiv.org/abs/1908.04722}{{\ttfamily 1908.04722}}].

\bibitem{DVN/4DEJQM_2020}
M.~Malte, S.~Bein and J.~Sonneveld, \emph{{Re-implementation of a search for
  supersymmetry in the $H_T$/missing $H_T$ channel (137 fb$^{-1}$;
  CMS-SUSY-19-006)}}, .

\bibitem{Bierlich:2022pfr}
C.~Bierlich et~al., \emph{{A comprehensive guide to the physics and usage of
  PYTHIA 8.3}},
  \href{https://doi.org/10.21468/SciPostPhysCodeb.8}{\emph{SciPost Phys.
  Codeb.} {\bfseries 2022} (2022) 8}
  [\href{https://arxiv.org/abs/2203.11601}{{\ttfamily 2203.11601}}].

\bibitem{Read:2002hq}
A.L.~Read, \emph{{Presentation of search results: The $CL_s$ technique}},
  \href{https://doi.org/10.1088/0954-3899/28/10/313}{\emph{J. Phys. G}
  {\bfseries 28} (2002) 2693}.

\bibitem{Fowlie:2024dgj}
A.~Fowlie, \emph{{The Bayes factor surface for searches for new physics}},
  \href{https://doi.org/10.1140/epjc/s10052-024-12792-9}{\emph{Eur. Phys. J. C}
  {\bfseries 84} (2024) 426}
  [\href{https://arxiv.org/abs/2401.11710}{{\ttfamily 2401.11710}}].

\bibitem{Crivellin:2023xbu}
A.~Crivellin, M.~Kirk and A.~Thapa, \emph{{Minimal model for the $W$-boson
  mass, $(g-2)_\mu$, $h\to\mu^+\mu^-$ and quark-mixing-matrix unitarity}},
  \href{https://doi.org/10.1103/PhysRevD.108.L031702}{\emph{Phys. Rev. D}
  {\bfseries 108} (2023) L031702}
  [\href{https://arxiv.org/abs/2305.03081}{{\ttfamily 2305.03081}}].

\bibitem{ATLAS:2021yij}
{\scshape ATLAS} collaboration, \emph{{Search for new phenomena in final states
  with $b$-jets and missing transverse momentum in $\sqrt{s}=13$ TeV $pp$
  collisions with the ATLAS detector}},
  \href{https://doi.org/10.1007/JHEP05(2021)093}{\emph{J. High Energy Phys.}
  {\bfseries 05} (2021) 093}
  [\href{https://arxiv.org/abs/2101.12527}{{\ttfamily 2101.12527}}].

\end{thebibliography}\endgroup
